\documentclass[twocolumn,times]{aastex62}

\graphicspath{{./}{figures/}}

\received{ \today}
\revised{} 
\accepted{} 
\submitjournal{ApJL}

\shorttitle{Spin evolution of MSP}
\shortauthors{Lan et al.}

\begin{document}

\title{Effect of irradiation on the spin of millisecond pulsars}

\email{lanshunyi@ynao.ac.cn}
\email{xiangcunmeng@ynao.ac.cn}


\author[0009-0001-8816-8523]{Shun-Yi Lan}
\affiliation{Yunnan Observatories, Chinese Academy of Sciences, Kunming 650011, PR China}
\affiliation{University of the Chinese Academy of Sciences, 19A Yuquan Road, Shijingshan District, Beijing 100049, PR China}

\author[0000-0001-5316-2298]{Xiang-Cun Meng}
\affiliation{Yunnan Observatories, Chinese Academy of Sciences, Kunming 650011, PR China}
\affiliation{Key Laboratory for the Structure and Evolution of Celestial Objects, Chinese Academy of Sciences, Kunming 650216, PR China}
\affiliation{International Centre of Supernovae, Yunnan Key Laboratory, Kunming 650216, PR China}




\begin{abstract}

A millisecond pulsar (MSP) is an old neutron star (NS) that has accreted material from its companion star, causing it to spin up, which is known as the recycling scenario. During the mass transfer phase, the system manifests itself as an X-ray binary. PSR J1402+13 is an MSP with a spin period of $5.89~{\rm ms}$ and a spin period derivative of $\log\dot{P}_{\rm spin}=-16.32$. These properties make it a notable object within the pulsar population, as MSPs typically exhibit low spin period derivatives. In this paper, we aim to explain how an MSP can posses high spin period derivative by binary evolution. By utilizing the stellar evolution code \textsc{MESA}, we examine the effects of irradiation on the companion star and the propeller effect on the NS during binary evolution. We demonstrate that irradiation can modify the spin period and mass of an MSP, resulting in a higher spin period derivative. These results suggest that the irradiation effect may serve as a key factor in explaining MSPs with high spin period derivatives.

\end{abstract}



\section{Introduction} \label{sec1}

A neutron star (NS) in a binary system can accretes material from its companion star and spins itself up to form a millisecond pulsar (MSP). This process is known as the recycling scenario. During the mass transfer phase, depending on the mass of the companion star, the system can manifests itself as a high-mass X-ray binary ($M_2 \gtrsim 10M_\odot$) or a low-mass X-ray binary (LMXB, $M_2 \lesssim 1.5M_\odot$). For detailed reviews on compact object binary evolution, references such as \cite{1991PhR...203....1B}, \cite{2002ApJ...565.1107P}, and \cite{2006csxs.book..623T} can be consulted. Presently, more than 3300 pulsars have been detected, including approximately 530 MSPs \citep{2005AJ....129.1993M}\footnote{The current population of pulsars can be viewed at \href{https://www.atnf.csiro.au/research/pulsar/psrcat/}{https://www.atnf.csiro.au/research/pulsar/psrcat/}}. Some of these MSPs are referred to as transitional MSPs  (tMSP) \citep{2009Sci...324.1411A,2013Natur.501..517P,2014MNRAS.441.1825B}, and there are also accreting millisecond X-ray pulsars (AMXPs) \citep{2021ASSL..461..143P}. TMSPs exhibit changes between rotation-powered and accretion-powered states, demonstrating a direct connection between radio MSPs and AMXPs. These observations strongly support the recycling scenario. For a recent review about tMSPs, one can refer \cite{2022ASSL..465..157P}. In comparison to normal pulsars, MSPs typically have spin periods lower than $30~{\rm ms}$, lower spin period derivatives $\log \dot P_{\rm spin}\sim-20$, and lower magnetic fields $B\sim 10^8~{\rm G}$. Most NSs can be born with strong magnetic fields $B\sim 10^{13.25}~{\rm G}$ \citep{2010MNRAS.401.2675P}, nevertheless, during the recycling process, the NS accretes material, leading to a reduction in its magnetic field \citep{1989Natur.342..656S}. Most observational results also support canonical evolutionary theory \citep[e.g.,][]{2012MNRAS.425.1601T}. However, there are exceptions, e.g., the MSP PSR J1402+13 \citep{2022ApJS..260...53A} with a spin period of $5.89~{\rm ms}$ and a spin period derivative of $\log \dot P_{\rm spin}=-16.32$.

The spin evolution of MSPs has always been an important aspect of study. It can help determine the equation of state of NS and constrain X-ray binary evolution theory. Based on the mass transfer rate value, as well as the spin period and magnetic field of the NS, the NS can experience either spin-up or spin-down evolution as a result. There are essentially three regimes depending on the mass transfer rate value: radio spin-down, propeller spin-down, and accretion spin-up \citep[e.g.,][]{2012MNRAS.425.1601T,2021MNRAS.502L..45B,2021ApJ...922..158L}. During the accretion spin-up regime, the NS accretes material transferred from its companion star and spins itself up. When the mass transfer rate decreases to a specific value, the propeller effect is activated. During the propeller regime, the transferred mass is expelled from the binary, preventing it from being accreted by the NS. The NS loses its angular momentum which is carried by ejected matter and the magnetic field. \citep{2018NewA...62...94R}. As the mass transfer rate value further decreases and even stop, the binary system will eventually host a radio pulsar.

Despite the progress made in understanding the overall perspective of the recycling scenario, uncertainties still remain. For instance, a birthrate problem emerged when \cite{1988ApJ...335..755K} proposed the birthrate of MSPs and the number of LMXBs, from which they descend in this framework, do not match. The irradiation effect in LMXBs could be a solution to the birthrate problem. Irradiation primarily affects stars with a convective envelope, which explains why it can have a significant impact on the evolution of LMXBs \citep{1991Natur.350..136P}. The irradiation-induced mass transfer cycles lead to a reduced time spent in the LMXB phase \citep{2000A&A...360..969R,2004A&A...423..281B,2008NewAR..51..869R}, which demonstrates the potential of the irradiation effect in addressing the birthrate problem. Additionally, the irradiation effect can explain the existence of certain MSPs with a positive orbital period derivative \citep{2021ASSL..461..143P}. A series of works by \cite{2012ApJ...753L..33B,2014ApJ...786L...7B,2015ApJ...798...44B,2017A&A...598A..35B} have found that the irradiation effect may play a crucial role in the formation of black widow/redback pulsars, with black widows possibly evolving from redbacks. The irradiation effect may play a key role in the formation of AMXPs such as SAX J1808.4-3658 \citep{2018MNRAS.479..817T,2020MNRAS.495..796G}. Irradiated systems may also serve as good progenitors for massive MSPs \citep{2020MNRAS.495.2509E}. In this paper, we will show that the irradiation effect is also a key factor to form the MSPs with high spin period derivative.

This paper is organised as follows: Section~\ref{sec2} describes our input physics and methods. Section~\ref{sec3} shows our results. In Section~\ref{sec4}, we discuss the possible indications about our results and give some conclusions. 

\section{Methods} \label{sec2}

We use the stellar evolution code Modules for Experiments in Stellar Astrophysics \citep[\textsc{MESA},version 10398;][]{2011ApJS..192....3P,2013ApJS..208....4P,2015ApJS..220...15P,2018ApJS..234...34P,2019ApJS..243...10P} to simulate possible evolution of LMXB and the spin of MSP. We assume that the NS has a canonical initial mass of $1.4~M_\odot$. Also, the binary evolution begin with a NS and a zero-age main-sequence companion star in a circular and synchronized orbit. The metallicity of the companion star is $Z=0.02$. The maximum age of the system is set to be $14~\rm{Gyr}$.

Mass transfer rate value is described by the Ritter scheme \citep{1988A&A...202...93R}.  We adopt the isotropic re-emission model \citep{2006csxs.book..623T} to compute the mass-loss from the binary system with the following parameters: $\alpha=0,~\beta=0.7,~\delta=0$, which correspond to the fraction of mass lost from the vicinity of companion star, the vicinity of NS and the circumbinary co-plane toroid, respectively. The mass accretion rate onto NS is $\dot{M}_{\rm NS}=(1-\alpha-\beta-\delta)|\dot{M_2}|$, where $\dot{M}_2$ is the mass lost rate of companion star. The $\dot{M}_{\rm NS}$ is limited by Eddington accretion rate \citep{2012MNRAS.425.1601T}
    \begin{equation}
        \dot{M}_{\rm Edd} \simeq3.0\times10^{-8}M_\odot\mathrm{yr}^{-1}~ \left(\frac{R_{\rm NS}}{13~{\rm km}} \right)\left(\frac{1.3}{1+X}\right),
    \end{equation}
where the $R_{\rm NS}=15~(M_{\rm NS}/M_\odot)^{1/3}~{\rm km}$ is the radius of the NS, and the $X$ is the fraction of hydrogen of transferred mass.

\subsection{Orbital evolution}
\label{sec2.1}
In our simulation, orbital angular momentum loss is dominated by three mechanisms: gravitational wave radiation, mass loss from the system and magnetic braking. 

The orbital angular momentum loss by gravitational wave radiation is calculated by \citep{1971ctf..book.....L}
    \begin{equation}
        {\dot{J}}_{\mathrm{GW}}=-{\frac{32G^{7/2}M_{\mathrm{NS}}^{2}M_{\mathrm{2}}^{2}(M_{\mathrm{NS}}+M_{\mathrm{2}})^{1/2}}{5c^{5}a^{7/2}}},
    \end{equation}
where $G$ is the gravitational constant, $c$ is the speed of light in vacuum and $a$ is the semi-major axis of the orbit. 

Material lost from the system will carry specific orbital angular momentum of NS away from system. We calculate the angular momentum lossed by this mechanism as
    \begin{equation}
        \dot{J}_{\text{ML}}=(\alpha+\beta+\delta)|\dot{M}_{\text{2}}|\bigg(\frac{M_{\text{NS}}}{M_{\text{NS}}+M_{\text{2}}}\bigg)^2\frac{2\pi a^2}{P_{\text{orb}}},
    \end{equation}
where the $P_{\text{orb}}$ is the orbital period. 

The magnetic braking will also cause the loss of orbital angular momentum. We use the standard magnetic braking prescription  \citep{1983ApJ...275..713R}:
    \begin{equation}
        \dot{J}_{\text{MB}}=-3.8\times10^{-30}M_{\text{2}}R_{\text{2}}^{\gamma}\Omega^3\text{dyn cm},
    \end{equation}
where $R_2$ is the radius of the companion star, $\gamma$ is the magnetic braking index in which we adopt the value $\gamma=4$, $\Omega$ is the spin angular velocity of the companion star.

    \subsection{Irradiation on companion star}
    \label{sec2.2}
During mass transfer phase, the system will manifest itself as an X-ray source. In this context, the companion star can be irradiated by X-rays. Such irradiation may significantly alter the evolution of LMXBs. For how irradiation affect the secular evolution of LMXBs, one can refer outstanding previous works, e.g., \cite{1991Natur.350..136P}, \cite{1994ApJ...434..283H}, \cite{1994A&A...291..842V}, \cite{2004A&A...423..281B}.

The X-ray irradiation effect on companion star is included in this work. The accretion luminosity is
    \begin{equation}
        L_{\rm X}=\frac{GM_{\rm NS}\dot{M}_{\rm NS}}{R_{\rm NS}},
    \end{equation}
follow \cite{1994ApJ...434..283H}, we calculate irradiation luminosity as
    \begin{equation}
 \\
L_{\rm irr} =\left\{\begin{array}{l l}{{\eta L_{\rm X}\left(\frac{R_{2}}{2a}\right)^{2}}}&{{\dot{M}_{2}<\dot{M}_{\rm Edd}}}\\ {{\eta L_{\rm X}\left(\frac{R_{2}}{2a}\right)^{2}\exp\left(1-\frac{|\dot{M}_{2}|}{\dot{M}_{\rm Edd}}\right)}}&{{\dot{M}_{2}\geq\dot{M}_{\rm Edd}}}\end{array}\right., 
    \end{equation}
where $\eta$ is the irradiation efficiency. The energy will deposit in the outer layer of companion star and the deposited energy is decrease with $e^{-\tau}$, where $\tau$ is the optical depth at a given radius.
    \subsection{Spin evolution of neutron star}
    \label{sec2.3}

The spin evolution of NS in binary system could be affected by multiple factors: gravitational wave radiation and magnetic radiation (dipole and multi-pole) of NS, spin-up and spin-down torques caused by the interactions between transferred material and magnetic field, etc. The gravitational wave and multi-pole radiation will not be discussed in this paper, since they may not dominant the energy loss \citep[e.g.,][]{2010ApJ...715..335K,2010ApJ...722..909P}.

We consider the accretion-induced magnetic filed decay of NS. Magnetic moment is calculated by following formula \citep{1989Natur.342..656S}:
    \begin{equation}
        \mu=\mu_0\left(1+\frac{\Delta M_{\rm NS}}{10^{-4}M_\odot}\right)^{-1},
    \end{equation}
where $\mu_0$ is the initial magnetic moment. $\Delta M_{\rm NS}$ is the accreted mass of NS.

Interactions between magnetic field and transferred mass are characterized by three radii: magnetosphere radius $r_{\rm mag}$, co-rotation radius $r_{\rm co}$ and light cylinder radius $r_{\rm lc}$. We calculate these radii by the following forms \citep{1979ApJ...234..296G,2018NewA...62...94R,2023ApJ...945....2Y}:
    \begin{equation}
        r_{\rm mag}\simeq1.5\times10^{8}\left(\frac{\mu}{\mu_0} \right)^{4/7}\left(\frac{M_{\rm NS}}{M_\odot} \right)^{-1/7}\left(\frac{\dot{M}_{\rm NS}}{\dot{M}_{\mathrm{Edd}} }\right)^{-2/7}\mathrm{cm},
    \end{equation}
    \begin{equation}
        r_{\rm co}\simeq1.5\times10^{8}\left(\frac{M_{\rm NS}}{M_\odot} \right)^{1/3}\left(\frac{P_{\rm spin}}{\rm 1~s} \right)^{2/3}\mathrm{cm},
    \end{equation}
    \begin{equation}
        r_{\rm lc}\simeq4.8\times10^9\frac{P_{\rm spin}}{\rm 1~s}~\text{cm}.
    \end{equation}
Follow \cite{2012Sci...335..561T}, the total torque acting on the NS is
    \begin{equation}
        N_{\mathrm{total}}=n\left(\dot{M}_{\rm NS}\sqrt{GM_{\rm NS}r_{\mathrm{mag}}}+\frac{\mu^2}{9r_{\mathrm{mag}}^3}\right)+N_\mathrm{radio},
    \end{equation}
where $n$ is a dimensionless parameter,
    \begin{equation}
        n=\tanh\left(\frac{1-(r_{{\rm mag}}/r_{{\rm co}})^{3/2}}{0.002}\right).
    \end{equation}
$N_\mathrm{radio}$ is the spin-down torque caused by magnetic dipole radiation. We calculate $N_\mathrm{radio}$ as \citep{2023ApJ...945....2Y}
    \begin{equation}
        N_\mathrm{radio}=-\frac{\mu^2}{r_{\mathrm{lc}}^3}.
    \end{equation}

We assume that the NSs in our simulations have an initial spin period $P_{\rm spin}=10~{\rm s}$ and a constant moment of inertia $I=10^{45}{\mathrm{g}}~{\mathrm{cm}}^{2}$. We also assume that all the transferred material is ejected from the magnetosphere during the propeller regime.

\section{Results} \label{sec3}

\begin{figure}
    \centering
    \resizebox{\hsize}{!}{\includegraphics{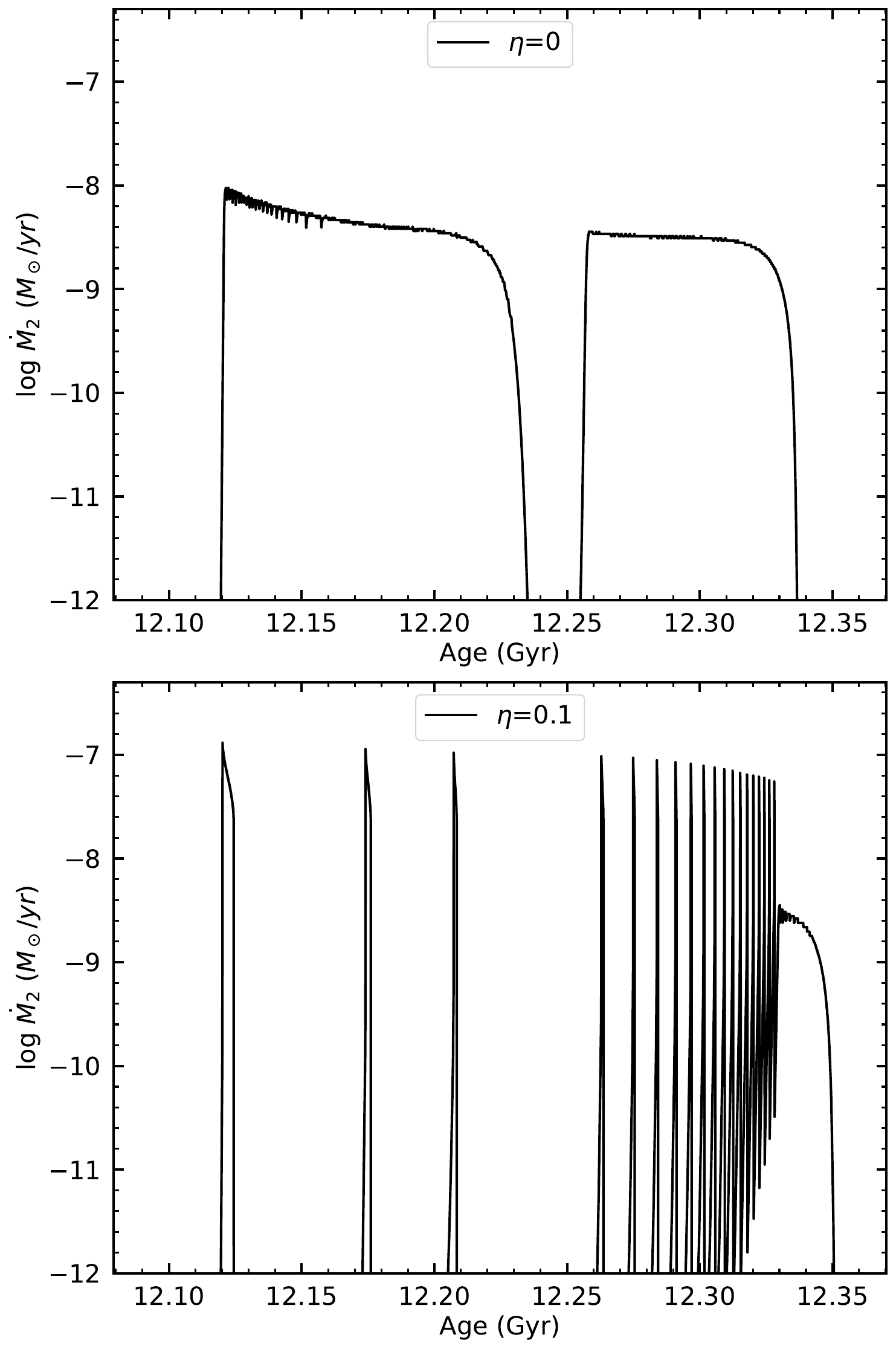}}
    \caption{Evolution of mass transfer rate value. The initial mass of companion star is $M_2=1.0~M_\odot$, initial orbital period is $P_{\rm orb}=5.0~\rm days$. Irradiation efficiencies are $\eta=0$ (top) and $\eta=0.1$ (bottom).}
    \label{fig1}
\end{figure}

We calculated a series of models of binary systems with an initial companion star mass $M_2=1.0~M_\odot$ and initial orbital periods $P_{\rm orb}=3.0,~5.0,10.0,~20.0~\rm{days}$ for different values of the irradiation efficiencies $\eta=0,~0.001,~0.01,~0.1$ and initial magnetic field $\mu_0$ range from $10^{29}~{\rm G~cm^3}$ to $10^{32}~~{\rm G~cm^3}$. Here, for simplicity, we only show the results from two examples with $\eta=0,~0.1$, both with $P_{\rm orb}=5.0~\rm{days}$ and $\mu_0=4.0\times10^{31}~{\rm G~cm^3}$, and will give a simple discussion on the effect of other parameters in Section~\ref{sec4}. Figure~\ref{fig1} shows the evolution of mass transfer rate value for different $\eta$. For $\eta=0$, the mass transfer occurs in a cyclic way instead in a continuous form (as in the case of $\eta=0$). During the initial phase of Roche-lobe overflow (RLO), the mass transfer rate value $\dot M_2$ increases, resulting in a corresponding increase in X-ray luminosity ($L_{\rm X}$). Irradiation causes the companion star to expand, leading to a higher mass transfer rate value $\dot M_2$ compared to the system without irradiation. Detachment of the companion star from its Roche lobe occurs when the expansion due to irradiation is outweighed by the thermal relaxation of the companion star. Orbital angular momentum loss subsequently induces the companion star to reattach, initiating a new cycle of mass transfer, and so forth. Further insights about the mechanism of the mass transfer cycle can be found in \cite{2000A&A...360..969R}. This phenomenon causes the system to constantly switch between the LMXB phase and the radio pulsar phase, which is quite different from that with $\eta=0$.

\begin{figure}
    \centering
    \resizebox{\hsize}{!}{\includegraphics{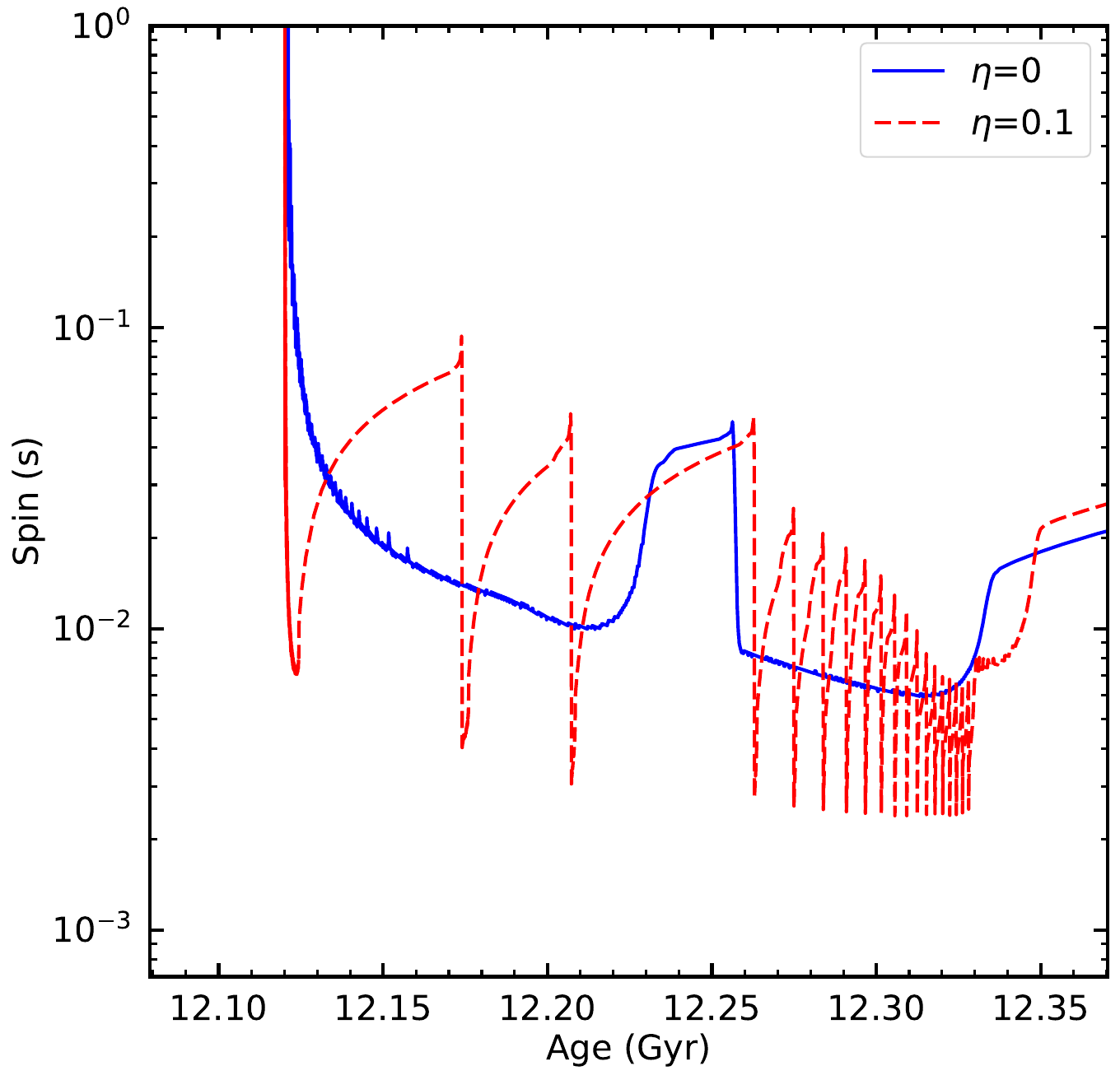}}
    \caption{Evolution of spin period corresponding to Figure~\ref{fig1}. Blue solid and red dashed lines indicate irradiation efficiencies $\eta=0$ and $0.1$, respectively. The initial values of magnetic fields for both evolutions are $\mu_0=4.0\times10^{31}~{\rm G~cm^3}$.}
    \label{fig2}
\end{figure}

\begin{figure}
    \centering
    \resizebox{\hsize}{!}{\includegraphics{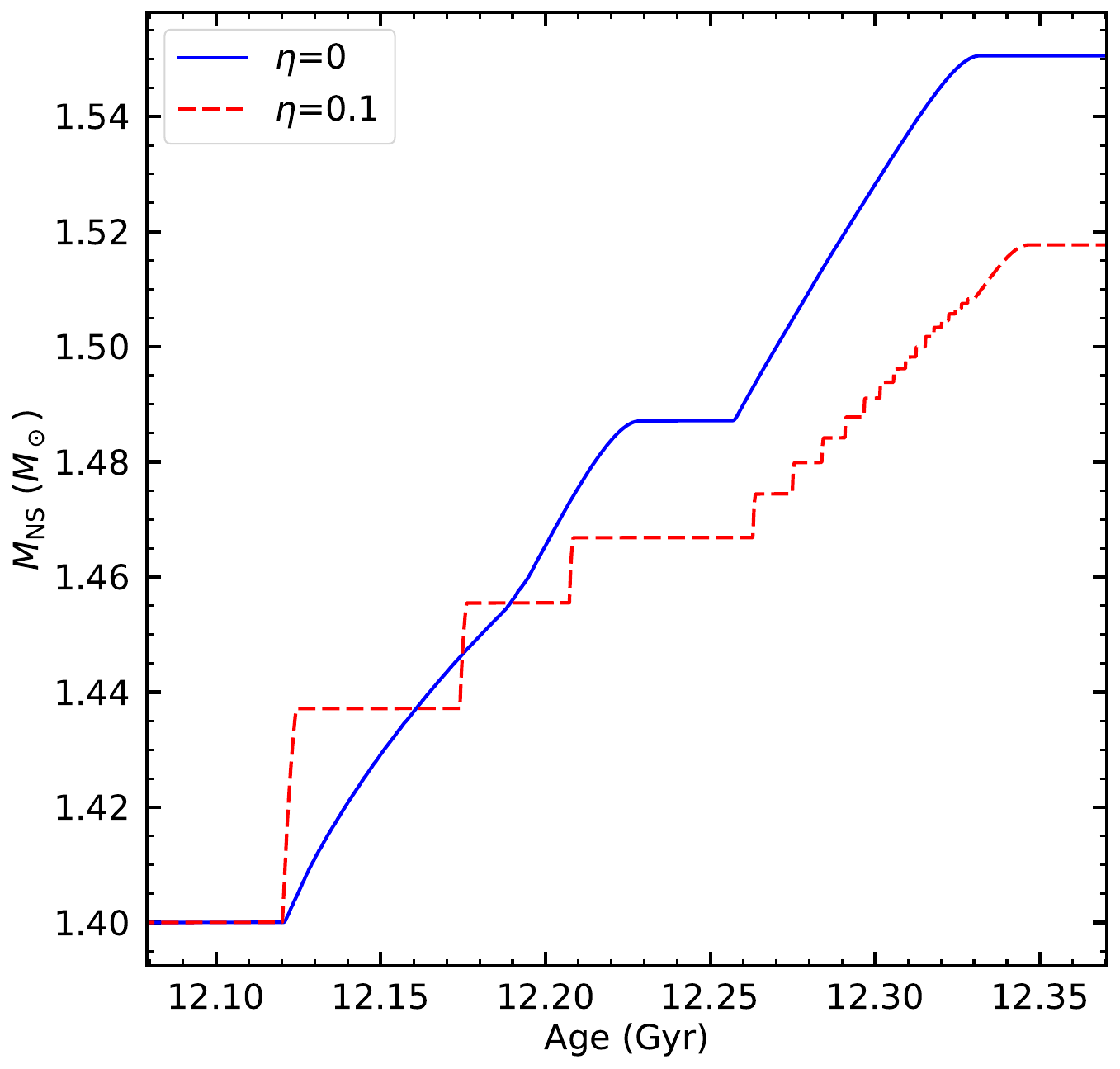}}
    \caption{Evolution of the mass of NS corresponding to Figure~\ref{fig1}. Blue solid and green dashed lines indicate irradiation efficiencies $\eta=0$ and $0.1$, respectively. }
    \label{fig3}
\end{figure}

\begin{figure}
    \centering
    \resizebox{\hsize}{!}{\includegraphics{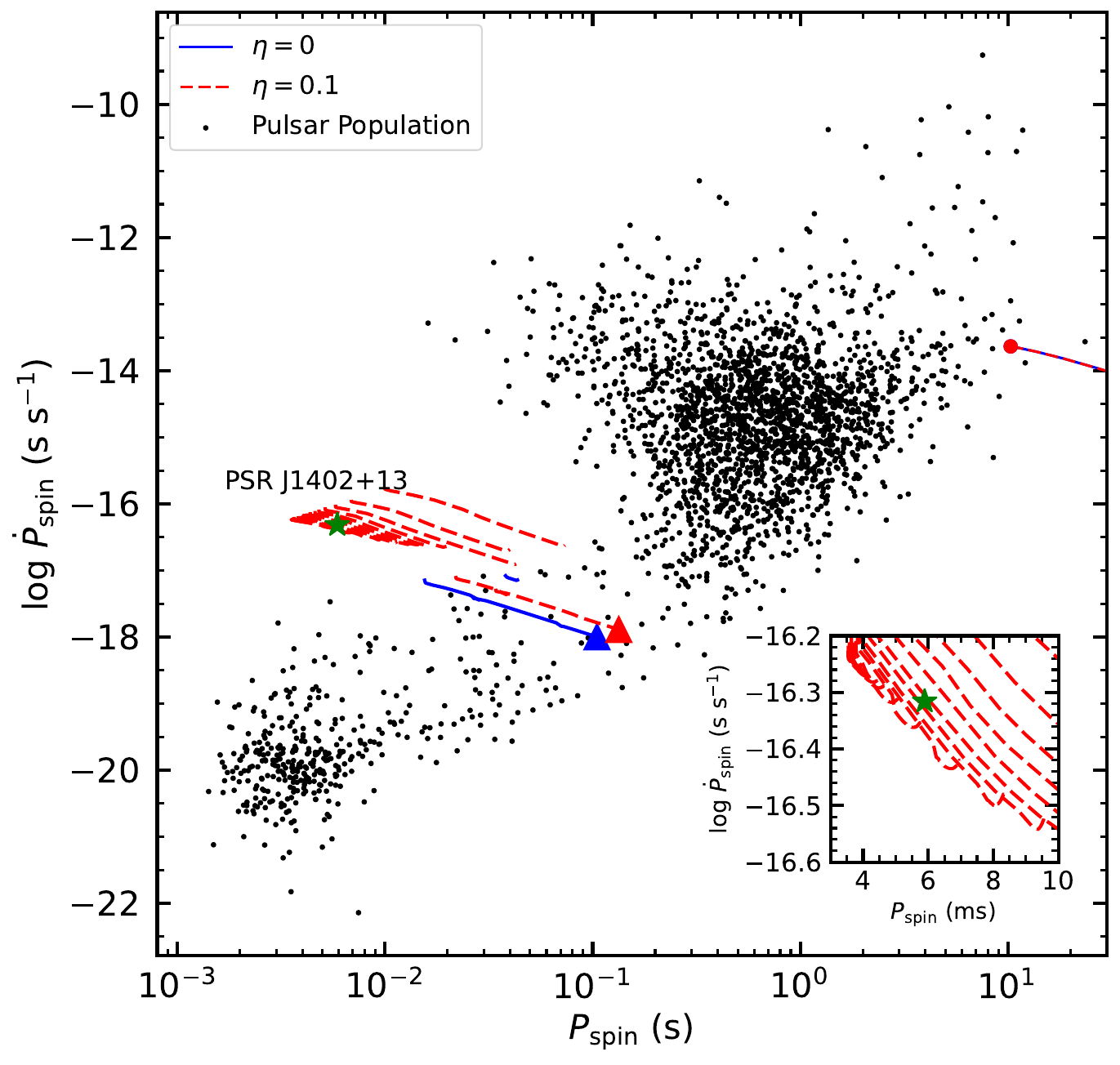}}
    \caption{Evolution tracks of radio phases in $P_{\rm spin} - \dot P_{\rm spin}$ diagram, corresponding to Figure~\ref{fig1}. The blue solid and red dashed lines are indicate $\eta=0$ and $0.1$, respectively. All line segments indicate $r_{\rm mag}>r_{\rm lc}$. The initial values of magnetic fields for both evolutions are $\mu_0=4.0\times10^{31}~{\rm G~cm^3}$. Solid circles and triangles are the evolution start and end points, respectively. Green solid star present the PSR J14002+13. Black points are current pulsar population, and data are from \href{https://www.atnf.csiro.au/research/pulsar/psrcat/}{https://www.atnf.csiro.au/research/pulsar/psrcat/}.}
    \label{fig4}
\end{figure}

Figure~\ref{fig2} shows the spin period evolution. The spin period evolution  of NS in an irradiated system is significantly different from that of none irradiated system. For non-irradiated system, i.e., $\eta=0$, for the most part the first and the second RLO, the NS is actually in a spin-equilibrium state {(i.e., $r_{\rm mag}\simeq r_{\rm co}$)} \citep[see also][]{2012MNRAS.425.1601T}, and constantly experiences minor propeller regimes during $\sim12.12-12.21~\rm{Gyr}$ and $\sim12.25-12.32~\rm{Gyr}$. On the other hand, for $\eta=0.1$, irradiation-induced mass transfer cycles cause the NS to experience multiple propeller regimes and radio spin-down phases before the mass transfer fully stops. Due to a higher mass transfer rate value, the NS can reach a lower spin period. 

Figure~\ref{fig3} shows the evolution of the mass of the NS. For $\eta=0.1$, although the mass transfer rate value during RLO is larger than that in the case of $\eta=0$, the total duration of RLO for the irradiated system is shorter than that for the non-irradiated system. Moreover, the accretion rate remains constrained by the Eddington limit. As a result, the final mass of the MSP in the irradiated system is smaller than that in the non-irradiated system.

In Figure~\ref{fig4}, we show the evolutionary tracks when binary systems host an radio pulsar in {$P_{\rm spin} - \dot P_{\rm spin}$} diagram. Same as previous figures, different line styles and colors indicate different values of irradiation efficiency. The line segments in Figure~\ref{fig4} present the evolutionary path segments corresponding to the case $r_{\rm mag}>r_{\rm lc}$, i.e., the system hosts a radio MSP. Each line segment evolves from the upper left to the lower right.  For $\eta=0$, due to interaction between magnetic field and transferred material, it is difficult for NS to possessing lower spin period and higher spin period derivative at same time during radio phases. For $\eta=0.1$, the tracks can cover a larger part of $P_{\rm spin} - \dot P_{\rm spin}$ diagram. Due to the irradiation-induced mass transfer cycles, the NS can reach a lower spin period and undergo several radio phases before the mass transfer completely stops. This enables the NS to possess both a lower spin period and higher spin period derivative during the radio phases. As we can see in Figure~\ref{fig4}, our calculations show that the irradiated models can explain the high value of the spin period derivative of PSR J1402+13.

\section{Discussion and Conclusions} \label{sec4}

In this work, we showed binary evolution calculations of a system composed of a canonical NS whose companion is a normal star with initial mass of $M_2=1.0~M_\sun$ in an orbit with initial period of $P_{\rm{orb}}=5~\rm{days}$. We considered both the irradiation effect and the propeller effect, tracking the spin period evolution of MSPs. Our goal was to provide an explanation for the formation of MSPs having high spin period derivatives.

In our simulations, the two most crucial free parameters affecting the spin evolution are the irradiation efficiency $\eta$ and the initial magnetic field $\mu_0$. Typically, $\eta$ is believed to range from 0.01 to 0.1 in LMXBs \citep{2004A&A...423..281B}, while population synthesis suggests that most NSs may have an initial magnetic field of $\mu_0\sim 10^{31}~{\rm G~cm^3}$ \citep{2010MNRAS.401.2675P}. The value of $\eta$ affects the peak mass transfer rate value during RLO and the number of mass transfer cycles \citep[e.g.,][]{2014ApJ...786L...7B}, which in turn determines the minimum spin period that an MSP can reach. Our additional tests (not shown here) suggest that NSs in binary systems with initial magnetic fields of $\mu_0\sim 10^{31}-10^{32}~{\rm G~cm^3}$ may produce radio MSPs with high spin period derivatives. The exact relationship between irradiation efficiency, the initial magnetic field of the NS, the initial orbital period, the initial mass of companion star, and the spin period and spin period derivative of MSPs varies complicatedly and still needs to be investigated in details in the future.

Due to the absence of other estimated parameters, we are currently unable to rigorously constrain the evolutionary history of PSR J1402+13\footnote{Although a catalog \href{http://astro.phys.wvu.edu/GalacticMSPs/GalacticMSPs.txt}{GalacticMSPs.txt} says an orbital period of 52.49 days, this parameter is absent from the ANTF catalog. We simply assume that the system is a binary, but have not incorporated this value into our study, although, our models can successfully reproduce this orbital period.}. In the future, we intend to conduct more comprehensive investigations utilizing binary population synthesis, encompassing topics such as but not limited to the progenitor system of PSR J1402+13 and the birth rate of MSPs having high spin period derivatives.

As shown in Figure~\ref{fig2} and Figure~\ref{fig3}, the final spin period and mass of the MSP can be altered by the irradiation effect. Furthermore, the irradiation effect can modify the evolution track on the $P_{\rm spin} - \dot P_{\rm spin}$ diagram, leading to a wider range of possible tracks. In conclusion, the canonical evolutionary theory faces difficulties in producing radio MSPs with high spin period derivatives, but the irradiation effect provides an potential way to conquer this problem.


\section*{Acknowledgments}

\begin{acknowledgments}
We are grateful to the anonymous referee for the constructive comments that helped us to improve the manuscript. S.L. would like to express sincere gratitude to Hai-Liang Chen, Zheng-Wei Liu, Zhen-Wei Li, Ph.Podsiadlowski, O.G.Benvenuto and A.J.Goodwin for their valuable helps and suggestions. This work is supported by the National Science Foundation of China and National Key R\&D Program of China (No. 2021YFA1600403), the National Natural Science Foundation of China (Nos. 11973080, 12333008 and 12288102). X.M. acknowledges support from the Yunnan Ten Thousand Talents Plan - Young \& Elite Talents Project, and the CAS `Light of West China' Program. X.M. acknowledges support from International Centre of Supernovae, Yunnan Key Laboratory (No. 202302AN360001), the Yunnan Revitalization Talent Support Program-Science \& Technology Champion Project (NO. 202305AB350003), Yunnan Fundamental Research Projects (NO. 202201BC070003) and the science research grants from the China Manned Space Project.
\end{acknowledgments}


\end{document}